\documentclass[preprint,prd,aps,nofootinbib]{revtex4}
\usepackage{graphicx}
\usepackage{diagbox}
\usepackage{float}
\usepackage{mathrsfs}
\usepackage{color}
\usepackage{slashed}
\begin{document}
\title{Improved covariant analysis of $B_c^+ \to \chi_{c1}(nP)$ decays and implications for the nature of $\chi_{c1}(3872)$}
\author{Shi-Hang Zhang$^{1,2,3*}$,
 Jia-Yi Zheng$^{1,2,3}$\footnote{These authors contributed equally},
 Wei Li$^{4}$,
 Su-Yan Pei$^{5}$, Tianhong Wang$^6$, Zhi-Hui Wang$^{7,8}$,  Chao-Hsi Chang$^{9,10}$, 
Guo-Li Wang$^{1,2,3}$\footnote{Corresponding author}}
\affiliation{$^1$ Department of Physics, Hebei University, Baoding 071002, China\\
$^2$ Hebei Key Laboratory of High-precision Computation and
 Application of Quantum Field Theory, Baoding 071002, China\\
$^3$ Hebei Research Center of the Basic Discipline for Computational Physics, Baoding 071002, China\\
$^4$ College of Science, Hebei Agriculture University, Baoding 071001, China\\
$^5$ Department of Basic Industrial Education, Hebei Vocational University of Industry and Technology,
Shijiazhuang 050091, China\\
$^6$ School of Physics, Harbin Institute of Technology, Harbin 150001, China\\
$^7$Key Laboratory of Physics and Photoelectric Information Functional Materials, North Minzu University, Yinchuan 750021, China\\
$^8$School of Electrical and Information Engineering, North Minzu University, Yinchuan 750021, China\\
$^9$ Institute of Theoretical Physics, Chinese Academy of Science, Beijing 100190, China\\
$^{10}$ CCAST(World Laboratory), P.O. Box 8730, Beijing 100190, China}
\begin{abstract}
We study the weak decays $B_c^+ \to \chi_{c1}(nP) \ell^+ \nu_\ell$
and $B_c^+ \to \chi_{c1}(nP) X$ (n=1,2,3) within the Bethe–Salpeter formalism, treating $\chi_{c1}(3872)$ as the conventional $\chi_{c1}(2P)$ charmonium. We upgrade the covariant hadronic transition amplitude to consistently evaluate the final-state wave function in its rest frame, enabling a reliable description of large-recoil processes and sizable relativistic corrections pertinent to highly excited charmonium. Our results provide a novel platform to probe the internal structure of $\chi_{c1}(3872)$ via $B_c$ decays. While the LHCb search for $B_c^+ \to \chi_{c1}(3872) \pi^+$ yielded only an upper limit, we demonstrate that this is primarily due to insufficient luminosity, approximately 20 times the current $B_c$ data sample would be required for a definitive observation. In contrast, we identify the semileptonic mode $B_c^+ \to \chi_{c1}(3872) \mu^+ \nu_\mu$
​as a far more promising channel, which could become accessible with merely twice the existing data set. Our predictions offer concrete guidance for upcoming LHCb analyses and complement ongoing efforts to resolve the nature of $\chi_{c1}(3872)$.
\end{abstract}
\maketitle
\section{Introduction}

$\chi_{c1}(3872)$ was discovered by the Belle Collaboration in 2003 ~\cite{7}, marking a new milestone in the study of hadronic states. Since its discovery, research on this particle has remained a hot topic in hadron physics. To this day, various possible internal structures have been investigated, including a conventional charmonium state $\chi_{c1}(2P)$ \cite{ted,suzuki,ktchao0,achasov,achasov1}, a compact $c\bar{c}q\bar{q}$ tetraquark state \cite{maiani,Matheus,thuang,hxing,zhangal}, a $D\bar{D}^*$ molecular state \cite{Swanson,Gamermann,slzhu,Braaten,Albaladejo,fzpeng,HYun,slzhu2}, a $c\bar{c}g$ hybrid state \cite{fclose,liba}, and mixed states such as a mixture of charmonium and molecular \cite{Matheus1,Ortega,fkguo,Takizawa,dongy,Karliner,Padmanath,cmeng,zyzhou,kangxw,Miyake}, a mixture of molecular and tetraquark  \cite{Grinstein,Carducci}, and a mixture of hybrid and molecular  \cite{wchen}, among others. However, so far no consensus has been reached, and the internal structure of $\chi_{c1}(3872)$ remains controversial.

Among these possible structures of $\chi_{c1}(3872)$, the conventional charmonium $\chi_{c1}(2P)$ interpretation remains a competitive option. There are two main reasons against identifying $\chi_{c1}(3872)$ as the charmonium $\chi_{c1}(2P)$. First, the mass of $\chi_{c1}(3872)$ is several tens of MeV lower than that predicted by potential models for
$\chi_{c1}(2P)$; however, studies have shown that this puzzle can be explained by the corrections from virtual hadron loops or coupled-channel effects \cite{ktchao0,achasov,ted2,Ferretti,Ferretti2}. Second, the experimental ratio of $\mathcal{R}_{\psi(2S)/\psi(1S)}=\frac{\Gamma(\chi_{c1}(3872)\to \psi(2S)\gamma)}{\Gamma(\chi_{c1}(3872)\to \psi(1S)\gamma)}$ disagrees with the theoretical prediction, this is the primary reason against the
$\chi_{c1}(2P)$ assignment. Nevertheless, the situation has changed with improvements in experimental precision. Initially, the ratio given by BaBar collaboration in 2009 was $\mathcal{R}_{\psi(2S)/\psi(1S)}=3.4\pm 1.4$ \cite{ex2009}, and the 2022 Particle Data Group value was
$\mathcal{R}_{\psi(2S)/\psi(1S)}=2.6\pm0.6$ \cite{pdg1}, both larger than the theoretical values. For example, our theoretical calculation in 2024 gave a value of $\mathcal{R}_{\psi(2S)/\psi(1S)}=1.77$ for $\chi_{c1}(2P)$ \cite{peisy}, and at that time we pointed out that the theoretical value was smaller than the experimental one, leaving the question of whether $\chi_{c1}(3872)$ is $\chi_{c1}(2P)$ still open. However, a few months later, the LHCb experiment released results with higher precision, giving a value of $\mathcal{R}_{\psi(2S)/\psi(1S)}=1.67\pm0.21\pm0.12\pm0.04$ \cite{exlhcb}. This is in excellent agreement with our theoretical value, and also consistent with some charmonium predictions \cite{ted,Ferretti2,ted3,fazio,ktzhao,ydong,Badalian,zhong1}. Previously, these theoretical results had been taken as evidence against $\chi_{c1}(3872)$ being $\chi_{c1}(2P)$; yet it is now clear that the conventional $\chi_{c1}(2P)$ interpretation remains a strong competitor for $\chi_{c1}(3872)$.

Experimentally, various modes have been attempted to produce $\chi_{c1}(3872)$, such as direct production in colliders:
$p\bar{p}\to \chi_{c1}(3872)+anything$ \cite{pbpX}, $pp\to \chi_{c1}(3872)+anything$ \cite{ppX}
$e^+e^-\to \chi_{c1}(3872) \gamma$ \cite{r3872}, and $e^+e^-\to \chi_{c1}(3872) \omega$ \cite{o3872}.
It can also be produced via hadronic decays:
$B\to \chi_{c1}(3872)K$ \cite{7}, $B^0_s\to \chi_{c1}(3872)\phi$ \cite{B0s},
and $\Lambda^0_b\to \chi_{c1}(3872) pK^-$ \cite{lhc2019}.
Last year, LHCb corraboration also made the first attempt to detect $\chi_{c1}(3872)$ in $B_c$ decays. They searched for $B^+_c\to \chi_{c1}(3872) \pi^+$ and reported an upper limit \cite{11}:
\begin{eqnarray}\label{eq1}
\mathcal{R}_{\psi(2S)}^{\chi_{c1}(3872)} = \frac{\mathcal{B}_{B_c^+ \to \chi_{c1}(3872)\pi^+}}{\mathcal{B}_{B_c^+ \to \psi(2S)\pi^+}} \times \frac{\mathcal{B}_{\chi_{c1}(3872) \to J/\psi \pi^+\pi^-}}{\mathcal{B}_{\psi(2S) \to J/\psi \pi^+\pi^-}} < 0.05 \, (0.06).
\end{eqnarray}
Therefore, searching for the process of $B_c$ meson decaying into $\chi_{c1}(3872)$ remains a future experimental goal, as this process is important and constitutes one of the key ways to discriminate the internal structure of $\chi_{c1}(3872)$.

To facilitate better experimental searches, it is also necessary to provide relatively precise theoretical studies. Therefore, this paper focuses on the process $B_c\to \chi_{c1}(nP)$ ($n=1,2,3$), where $\chi_{c1}(3872)$ is treated as the $\chi_{c1}(2P)$. The main focus of our study is on relativistic corrections. It is generally believed that both the $B_c$ meson and charmonium are composed of heavy quarks, and thus relativistic corrections are not large. However, this is not actually the case, because our previous studies have found that the relativistic corrections for excited heavy meson states are much larger than expected. If a nonrelativistic model is adopted, it would introduce significant errors. The $\chi_{c1}(2P)$ is exactly such a case. For example, in our previous work (Ref. \cite{wgl1}), we calculated the mean square relative velocity of quarks inside $\eta_c$ $(J/\psi)$ to be $v^2=0.25$, while for $\chi_{c1}(2P)$ the value is $v^2=0.39$. This indicates that, despite being heavier in mass, the relativistic corrections for the $\chi_{c1}(2P)$ are much larger than those for $\eta_c$ $(J/\psi)$. Indeed, this is borne out in Ref. \cite{19}, where we calculated the relativistic corrections for the semileptonic decay of $B_c$ to charmonium. It was found that the relativistic effects for $B_c\to \chi_{c1}(1P)e\bar{\nu}_{e}$ reach $34\%$, while those for the
$B_c\to \chi_{c1}(2P)e\bar{\nu}_{e}$ state are $43\%$ (or $45\%$). It is thus clear that if $\chi_{c1}(3872)$ is a
$\chi_{c1}(2P)$ state, its relativistic effects must be carefully taken into account.

In this paper, we upgrade the formulas for the hadronic transition matrix elements from Ref. \cite{19}, recalculate the semileptonic decays of $B_c$ to charmonium, and supplement the calculations of nonleptonic decays. We also compute the ratios of the two, since according to Ref. \cite{qzhao}, these ratios are expected to be universal and could be reliably predicted in theory, thereby providing deeper insights into the nature of the $\chi_{c1}(3872)$. Compared with the old matrix element formulas, the formulas in this paper are more covariant and can better handle large-recoil processes. They can also more accurately compute processes with large relativistic corrections, which is particularly important for highly excited states such as $\chi_{c1}(2P)$ and $\chi_{c1}(3P)$.

From the perspective of the $B_c^+$ meson, studying its decays is also of great significance. Since its discovery experimentally in 1998 by the CDF Collaboration~\cite{1}, the $B_c^+$ meson has served as an important platform for testing quantum chromodynamics (QCD) and exploring weak interactions. Unlike charmonium or bottomonium, the two heavy quarks in the $B_c$ meson carry distinct and explicit flavor quantum numbers and cannot annihilate via strong or electromagnetic interactions; therefore, the $B_c$ meson can only decay through weak interactions. This feature endows the $B_c$ meson with a relatively long lifetime and a rich variety of decay modes, providing unique opportunities for both experimental and theoretical studies. In particular, the enormous amount of data accumulated by experiments such as LHCb has made it possible to perform detailed studies of the weak decay properties of the $B_c$ meson~\cite{2}.
Over the past few decades, various theoretical methods have been applied to study the weak decays of the
$ B_c$ to $\chi_{c1}(1P)$, e.g., Refs. \cite{me2002,Kiselev,EH,MAI,lv2,azizi,WANG,zr,zr2,rlzhu,jlu,hnli}, but only a few papers have investigated the transitions of $ B_c \to \chi_{c1}(2P)$ \cite{19,ymwang,DE,zhangzq1,zhangzq2,zhangzq3}. Therefore, it is also necessary to carefully study the decays of $ B_c \to \chi_{c1}(2P)$.

In this paper, the Bethe-Salpeter (BS) equation \cite{BS} method is adopted, which is the relativistic dynamical equation for describing mesons. Since the mesons under consideration are composed of two heavy quarks, the instantaneous approximation is a suitable choice. Therefore, we do not solve the full BS equation but instead its instantaneous version, namely the Salpeter equation \cite{Salpeter}. The Salpeter equation has broad applications in particle physics \cite{Brodsky,Thompson,Lucha,Colangelo,Resag,Munz,ktchao,Loringa}. We upgrade its solution by abandoning the construction of wave function representations based on partial waves (or $^{2S+1}L_{J}$), and instead construct wave function representations based on $J^P$ \cite{0-,1-,1++}, which are then substituted into the Salpeter equation for solution. In this way, we obtain wave functions containing more relativistic information \cite{wgl2}, and apply them to processes such as mass spectra \cite{spec,wgl2}, weak decays \cite{zj,fuhf,19,zhout}, and strong decays \cite{ann0-,wgl5,3930}, and electromagnetic processes~\cite{li01,pei02,li02,peisy}, achieving results in good agreement with experiments.

The paper is organized as follows. Section II presents the formulas for semi-leptonic and non-leptonic decays. Sec. III provides the formulae for the hadronic transition amplitudes. Sec. IV describes the relativistic wave functions of the mesons involved in this work. Sec. V presents our results and discussions. Finally, Sec. VI gives our conclusions.

\section{Semileptonic and Nonleptonic decay Formulas}
\begin{figure}[!htb]
\begin{minipage}[c]{1\textwidth}
\includegraphics[width=3in]{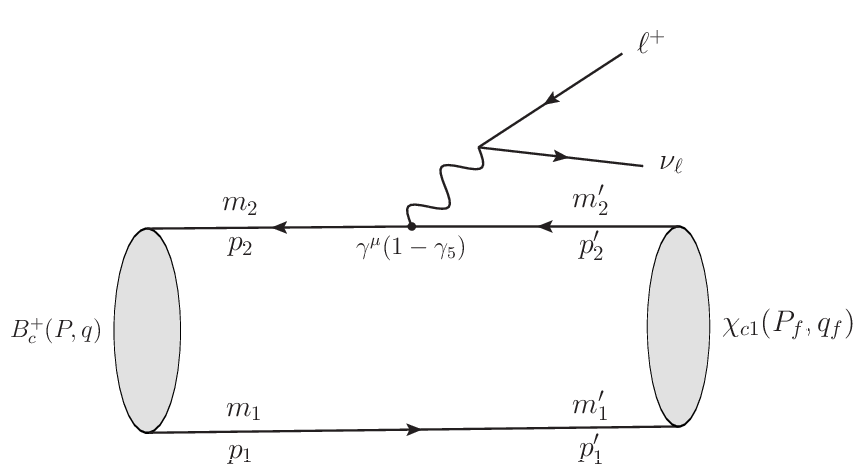}
\end{minipage}
\caption{Feynman diagram for  $B^{+}_c\rightarrow \chi_{c1}\ell^+\nu_{\ell}. $ }
\label{BQfeynman}
\end{figure}

The semileptonic decay $B_c^{+}\rightarrow \chi_{c1}\ell^+\nu_{\ell}$ is induced by the weak transition of $\bar{b}\to \bar{c}$, and the corresponding Feynman diagram is shown in Fig. \ref{BQfeynman}, the transition amplitude for the decay $B_c^{+}\rightarrow \chi_{c1}\ell^+\nu_{\ell}$ can be written as
\begin{eqnarray}
T=\frac{G_F}{\sqrt{2}}V_{cb}\bar{u}_{\nu_{\ell}}\gamma^{\mu}(1-\gamma_{5})v_{\ell}\langle{\chi_{c1}|J_\mu|B_c}\rangle,
\end{eqnarray}
where $G_F$ is the Fermi weak coupling constant, $V_{cb}$ is the CKM matrix element,  $u_{\nu_{\ell}}$ is the spinor of the neutrino $\nu_{\ell}$, $v_{\ell}$ is the spinor of the antilepton $\ell^+$, and $J_{\mu}$ is the charged weak current responsible for the $\bar b \to \bar c$ transition, etc.

Unlike the leptonic part $\bar{u}_{\nu_{\ell}}\gamma^{\mu}(1-\gamma_{5})
v_{\ell}$, the calculation of hadronic transition matrix element $\left \langle \chi_{c1}|J_\mu|B_c  \right \rangle$ is model-dependent. We will present our calculation method in detail in the next section; here we only give the definition of the form factors:
\begin{eqnarray}\label{transi}
\left \langle \chi_{c1}|J_\mu|B_c  \right \rangle =K_1~MM_f~\epsilon^*_{\mu }+K_2~\epsilon^*\!\cdot\!P P_\mu +K_3~\epsilon^*\!\cdot\!P P_{f\mu}  +i~K_4~\varepsilon_{ \mu\epsilon^* P P_f},
\end{eqnarray}
where $M$ and $P$ are the mass and momentum of $B_{c}$, $M_f$ and $P_f$ are the mass and momentum of $\chi_{c1}$, $\epsilon_{\mu }$ denotes the polarization vector of $\chi_{c1}$, $K_i$ ($i=1,2,3,4$) are the form factors, $\varepsilon_{\mu\nu\alpha\beta}$ is the Levi-Civita symbol, and we have used the following abbreviation: $\varepsilon^*_{ \mu\epsilon P P_f}\equiv\varepsilon^*_{ \mu\alpha\beta\gamma }\epsilon^\alpha P^\beta P_f^\gamma$.

After taking the modulus squared of the transition amplitude, averaging over the spins of the initial meson and summing over polarizations of final state, we obtain:
\begin{eqnarray}
\sum{|T|^2=\frac{G^2_F}{2}V^2_{cb}\ell^{\mu\nu}h_{\mu\nu}},
\end{eqnarray}
where $\ell^{\mu\nu}$ represents the leptonic tensor, which can be expressed as $\ell^{\mu \nu} \equiv {\sum}\bar{\mu}_{\nu_\ell}\gamma^\mu(1-\gamma_5)\upsilon _\ell\bar{\upsilon }_\ell(1+\gamma_5)\gamma^\nu\mu_{\nu_\ell}$, and $h_{\mu\nu}$ denotes the hadronic tensor, it is usually expressed as
\begin{eqnarray}
h_{\mu \nu} =&& {\sum}\left \langle B_c|J_\nu^+| \chi_{c1} \right \rangle \left \langle \chi_{c1}|J_\mu|B_c  \right \rangle\nonumber \\
=&&-\alpha g_{\mu \nu}+\beta_{++}\left(P+P_f\right)_{\mu}\left(P+P_f\right)_{\nu}+\beta_{+-}\left(P+P_f\right)_{\mu}\left(P-P_f\right)_{\nu} \nonumber \\
&&+\beta_{-+}\left(P-P_f\right)_{\mu}\left(P+P_f\right)_{\nu}+\beta_{--}\left(P-P_f\right)_{\mu}\left(P-P_f\right)_{\nu} \nonumber \\
&&+ i \gamma \varepsilon_{\mu \nu \rho \sigma}\left(P+P_f\right)^{\rho}\left(P-P_f\right)^{\sigma},
\end{eqnarray}
where the coefficients $\alpha$, $\beta_{\pm\pm}$ and $\gamma$ are functions of the form factors $K_i~ (i=1,2,3,4)$.

Thus, the differential decay rate of this exclusive process can be written as
\begin{eqnarray}
\frac{d^{2} \Gamma}{d x d y}= && \left|V_{cb}\right|^{2} \frac{G_{F}^{2} M^{5}}{32 \pi^{3}}\left\{\alpha \frac{\left(y-\frac{m_{\ell}^{2}}{M^{2}}\right)}{M^{2}}+2 \beta_{++} \right. \nonumber \\
&& \times{\left[2 x\left(1-\frac{M_f^{ 2}}{M^{2}}+y\right)-4 x^{2}-y+\frac{m_{\ell}^{2}}{4 M^{2}}\left(8 x+\frac{4 M_f^{2}-m_{\ell}^{2}}{M^{2}}-3 y\right)\right]} \nonumber \\
&& +\left(\beta_{+-}+\beta_{-+}\right) \frac{m_{\ell}^{2}}{M^{2}}\left(2-4 x+y-\frac{2 M_f^{2}-m_{\ell}^{2}}{M^{2}}\right)+ \beta_{--} \frac{m_{\ell}^{2}}{M^{2}}\left(y-\frac{m_{\ell}^{2}}{M^{2}}\right) \nonumber \\
&& \left.-\gamma\left[y\left(1-\frac{M_f^{2}}{M^{2}}-4 x+y\right)+\frac{m_{\ell}^{2}}{M^{2}}\left(1-\frac{M_f^{2}}{M^{2}}+y\right)\right]\right\},\nonumber\\
\end{eqnarray}
where {$m_\ell$ and $E_\ell$} are the mass and energy of the final charged lepton $\ell^+$, respectively; and $x\equiv E_\ell/M,~y\equiv (P-P_f)^2/M^2$.

\begin{figure}[!htb]
\begin{minipage}[c]{1\textwidth}
\includegraphics[width=3in]{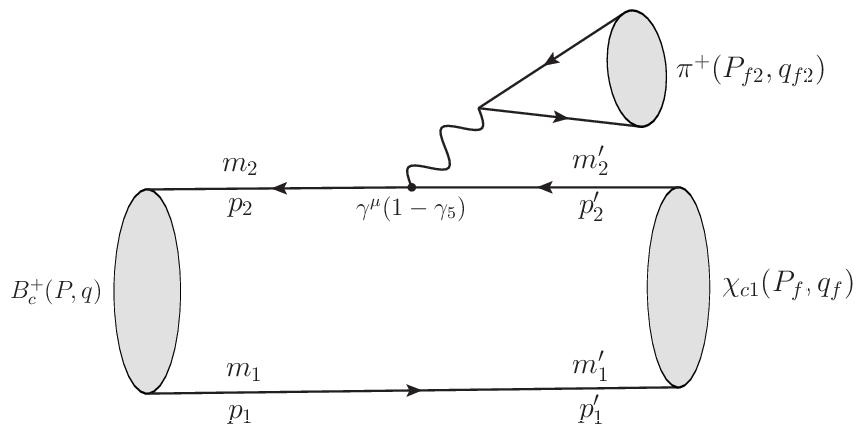}
\end{minipage}
\caption{Feynman diagram for  $B^{+}_c\rightarrow \chi_{c1} \pi^+. $ }
\label{Nfeynman}
\end{figure}

Besides the leptonic decays, we also evaluate a number of color-favored nonleptonic decays, $B_c^+ \rightarrow \chi_{c1} X$, where $X$ can be $\pi$, $K$, etc. Taking the case $X = \pi^+$ as an example, we present the corresponding Feynman diagram in Fig. \ref{Nfeynman}. Adopting the naive factorization approach, the transition amplitude for the nonleptonic decay can be expressed as the product of the hadronic transition matrix element and decay constant:
\begin{eqnarray}
T=\frac{G_F}{\sqrt{2}}V_{cb}V^*_{ud}a_1\langle{\chi_{c1}}|J_{\mu}|{B^+_c}
\rangle\langle{\pi^+}|J'^{\mu}|{0}\rangle,
\end{eqnarray}
where the $V_{ud}$ is also the CKM matrix element, $J'^{\mu}$ is the charged weak  current responsible for the $d \to u$ transition, $a_1=c_1+\frac{1}{N_c}c_2$, with $N_c$ indicating the number of color, and $c_1$ and $c_2$ being the Wilson coefficients. The matrix element $\langle{\pi^+}|J'^{\mu}|{0}\rangle$ is related to the decay constant, i.e., $\langle{\pi^+}|J'^{\mu}|{0}\rangle=if_{\pi}P^{\mu}_{\pi}$, where $f_{\pi}$ and $P^{\mu}_{\pi}$ ($=P^{\mu}_{f_2}$) are the decay constant and momentum of the $\pi$ meson. When $X$ corresponds to a vector meson, e.g., $K^{*+}$, the annihilation matrix element can be written as $\langle{K^{*+}}|J'^{\mu}|{0}\rangle=M_{K^{*+}}f_{K^{*+}}\epsilon^{\mu}_{K^{*+}}$, where $M_{K^{*+}}$ is the mass of $K^{*+}$, $f_{K^{*+}}$ denotes its decay constant, and $\epsilon^{\mu}_{K^{*+}}$ represents its polarization vector.

\section{Hadronic TRANSITION AMPLITUDE}
The calculation of the hadronic transition amplitude is model dependent.
In our method, according to the Mandelstam mechanism, it is expressed as an overlap integral over the wave functions of the initial and final mesons,
\begin{eqnarray}\label{transition matrix}
\left \langle \chi_{c1}|J_\mu|B_c^+  \right \rangle &&=\int\frac{d^4q}{(2\pi)^4}\frac{d^4q_{_f}}{(2\pi)^4}Tr[\bar{\chi}_{_{P_f}}(q_{_f})S^{-1}_1(p_{_1})\chi_{_P}(q)\gamma_\mu (1 - \gamma^5)(2\pi)^4\delta^4(p_{_1}-p'_{_1})] \nonumber\\
&&=\int\frac{d^4q}{(2\pi)^4}Tr[\bar{\chi}_{_{P_f}}(q_{_f})S^{-1}_1(p_{_1})\chi_{_P}(q)\gamma_\mu (1 - \gamma^5)],
\end{eqnarray}
where, $\chi_{_P}(q)$ is the BS wave function which is function of total momentum $P$ and internal relative momentum $q$ of the corresponding meson, with $\bar{\chi} = \gamma_0 \chi^\dagger \gamma_0$; $S_1$ denotes the propagator of the quark with momentum $p_1$. The relationship between the quark momentum and the internal relative momentum is given by:
$p_{_i}=\alpha_i P+J_iq$, $\alpha_i\equiv\frac{m_i}{m_1+m_2}$,
$i = 1, 2$ denote the quark and antiquark, respectively, with $m_1$ and $m_2$ being their masses, and $J_i = (-1)^{i+1}$. For the final meson, we have the similar formula: $p'_{i}=\alpha'_{i} P_f+J_iq_f$, $\alpha'_{i}\equiv\frac{m'_{i}}{m'_{1}+m'_{2}}$.

Since we solve the instantaneous version of the BS equation, namely, the Salpeter equation, instead of the full BS equation. Consequently, the above expression requires simplification. To proceed, we first provide a brief overview of both equations. The BS equation takes the form  \cite{BS}:
\begin{eqnarray}
\chi_{_{_P}}(q)=iS_1(p_{_1})\int\frac{d^4k}{(2\pi)^4}V(P,k,q)\chi_{_{_P}}(k)S_2(-p_{_2}),
\end{eqnarray}
$V(P, k, q)$ is the integral kernel between the quark and antiquark in the meson. Since the instantaneous approximation is well-suited for heavy mesons, Salpeter applied it to the BS equation and  reduced it to the Salpeter equation. The core idea of this approximation is to neglect the interaction propagation time between the quarks inside the meson. Thus, in the meson's center-of-mass system (CMS):
\begin{eqnarray}
V(P,k,q)\simeq V(\vec{k},\vec{q})=V(k_{_\bot},q_{_\bot}),
\end{eqnarray}
where $q_{_{\bot}}\equiv q-\frac{P\cdot q}{M}P$, and $q_{_{\bot}}=(0,\vec{q})$ in the CMS of $P$. In our calculations, the revised Cornell potential is employed, with its explicit expression provided in Appendix.

Define the following two functions:
\begin{eqnarray}
\varphi_{_P}(q_{_\bot})\equiv i \int\frac{dq_{_{_P}}}{2\pi}\chi_{_{_P}}(q),~~
\eta(q_{_\bot})\equiv\int\frac{d k_{_\bot}}{(2\pi)^3}V(k_{_\bot},q_{_\bot})\varphi_{_P}(k_{_\bot}),
\end{eqnarray}
where $\varphi_{_P}(q_{_\bot})$ denotes the Salpeter wave function. Thus, the BS equation becomes:
\begin{eqnarray}\label{bse}
&&\chi_{_P}(q)=S_1(p_{_1})\eta(q_{_\bot})S_2(-p_{_2}).
\end{eqnarray}

After integrating out the time component, the BS equation reduces to the Salpeter equation  \cite{Salpeter}:
\begin{eqnarray} \label{Salpeter}
(M - \omega_{1} - \omega_{2}) \varphi^{++}(q_{_\bot}) &=& \Lambda_{1}^{+}(q_{_\bot}) \eta(q_{_\bot}) \Lambda_{2}^{+}(q_{_\bot}), \nonumber\\
(M + \omega_{1} + \omega_{2}) \varphi^{--}(q_{_\bot}) &=& -\Lambda_{1 }^{-}(q_{_\bot})\eta(q_{_\bot}) \Lambda_{2}^{-}(q_{_\bot}), \\
\varphi^{+-}(q_{_\bot}) &=& \varphi^{-+}(q_{_\bot}) = 0,
\end{eqnarray}
where $\omega_{i} =\sqrt{m_{i}^{2}-p_{{{i\bot}}}^2}= \sqrt{m_{i}^{2}-q_{{_\bot}}^2}$ being the quark energy, and
\begin{eqnarray}\Lambda_{i}^{\pm}(q_{_\bot})&=&\Lambda^\pm_i(p_{i\bot})=\frac{1}{2\omega_{i}}\left[\frac{\slashed{P}}{M}\omega_{i}\pm(J_im_{i}+\slashed{p}_{i\bot})  \right] = \frac{1}{2 \omega_{i }}\left[\frac{\slashed P}{M} \omega_{i } \pm J_i(m_{i}+\slashed q_{{_\bot}})\right],\nonumber\\
\varphi^{\pm\pm}(q_{_\bot}) &\equiv& \Lambda_{1}^{\pm}(q_{_\bot}) \frac{\slashed P}{M} \varphi(q_{_\bot}) \frac{\slashed P}{M} \Lambda_{2 }^{\pm}(q_{_\bot}),~~
\varphi(q_{_\bot})=\varphi^{++}(q_{_\bot})+\varphi^{+-}(q_{_\bot})+\varphi^{-+}(q_{_\bot})+\varphi^{--}(q_{_\bot}).\nonumber
\end{eqnarray}

With these formulas, in Ref. \cite{wgl3}, we present the calculation formula $\mathcal{M}^A_{\mu}$ for the hadronic matrix element $\left \langle \chi_{c1}|J_\mu|B_c^+  \right \rangle$ appearing in Eq. (\ref{transition matrix}):
\begin{eqnarray}\label{ampA}
\mathcal{M}^A_{\mu}=\int\frac{d^3q_{_\bot}}{(2\pi)^3}Tr\left[
\bar{\varphi}^{++}(q_{_{f\bot}})\frac{\slashed{P}}{M}\varphi^{++}(q_{_\bot})\gamma_\mu (1 - \gamma^5)\right],
\end{eqnarray}
where the connection between the internal relative momenta of the initial and final mesons is expressed as $q_{_{f\bot}}=q_{_{\bot}}-\alpha'_1 P_{_{f\bot}}$, namely, $\vec{q}_{_f}=\vec{q}-\alpha'_1\vec{P_f}$ in the CMS of the $B_c$ meson. This correspondence, rooted in the spectator assumption, is a standard prescription in relativistic quark models \cite{isgur1,isgur2,DE}. Nevertheless, we note that the internal momentum of the final state is still parameterized in the CMS of  initial state. Considering that the meson wave functions are intrinsically obtained in their respective rest frames, such a treatment can lead to ambiguities, particularly in high-recoil processes. To address this, we need a modified amplitude expression wherein the final-state wave function is consistently evaluated in its rest frame.

Following the approach of Ref. \cite{zhout} and substituting Eq. (\ref{bse}) into Eq. (\ref{transition matrix}), the hadronic matrix element $\left \langle \chi_{c1}|J_\mu|B_c^+  \right \rangle$ is then expressed as,
\begin{eqnarray}\label{newa1}
&&\int\frac{d^4q}{(2\pi)^4}Tr\left[S_2(-p'_{_2})\bar{\eta}(q_{_{f\top}})S_1(p'_{_1})S_1^{-1}(p_{_1})S_1(p_{_1})\eta(q_{_\bot})S_2(-p_{_2})\gamma_\mu (1 - \gamma^5)]\right] \nonumber\\
&&=\int\frac{d^4q}{(2\pi)^4} Tr  \left[ S_2(-p'_{_2}) \frac{\slashed{P}_f}{M_f} \left(\widetilde{\Lambda}_2^+(p'_{_{2\top}})+\widetilde{\Lambda}_2^-(p'_{_{2\top}})\right) \bar{\eta}(q_{_{f\top}})
\left(\widetilde{\Lambda}_1^+(p'_{_{1\top}})+\widetilde{\Lambda}_1^-(p'_{_{1\top}})\right)\frac{\slashed{P}_f}{M_f}  \right. \nonumber\\ &&~~~~~~~~~~~~~~~~~~~~~\left.\times S_1(p_{_1})\eta(q_{_\bot})S_2(-p_{_2})\gamma_\mu (1 - \gamma^5)] \right]  \\
&&\simeq\int\frac{d^4q}{(2\pi)^4} Tr  \left[ S_2(-p'_{_2}) \frac{\slashed{P}_f}{M_f} \widetilde{\Lambda}_2^+(p'_{_{2\top}}) \bar{\eta}(q_{_{f\top}})\widetilde{\Lambda}_1^+(p'_{_{1\top}}) \frac{\slashed{P}_f}{M_f} S_1(p_{_1})\eta(q_{_\bot})S_2(-p_{_2})\gamma_\mu (1 - \gamma^5)] \right], \nonumber
\end{eqnarray}
where $q_{_{f\top}} \equiv q_{_f}-\frac{P_f\cdot q_{_f}}{M^2_f}P_f $ has been defined. As a spectator, $p'_{_1}=p_{_1}$. In the second equality above, we have applied the relation
\begin{eqnarray}1=\frac{\slashed{P}_f}{M_f}\frac{\slashed{P}_f}{M_f}=
\frac{\slashed{P}_f}{M_f} \left(\widetilde{\Lambda}_2^+(p'_{_{2\top}})+\widetilde{\Lambda}_2^-(p'_{_{2\top}})\right)=
\left(\widetilde{\Lambda}_1^+(p'_{_{1\top}})+\widetilde{\Lambda}_1^-(p'_{_{1\top}})\right)\frac{\slashed{P}_f}{M_f},\nonumber
\end{eqnarray}
where $
\widetilde{\Lambda}^\pm_i(p'_{_{i\top}})=\frac{1}{2\widetilde{\omega}_{i}}\left[\frac{\slashed{P}_{f}}{M_f}\widetilde{\omega}_{i}\pm(J_im'_{i}+\slashed{p'}_{i\top})  \right],
$ with $\widetilde{\omega}_{i}\equiv\sqrt{{m'_{i}}^{2}-{p'}_{_{i\top}}^2}=\sqrt{{m'_{i}}^2-q_{_{f\top}}^2}$.
In the third equality, we impose the condition that the positive-energy wave function dominates over its negative-energy counterpart.

Next, we express the propagator in terms of projection operators as follows:
\begin{eqnarray}
&&S_i(J_ip_{_i})=\frac{\Lambda_i^+(p_{_{i\bot}})}{p_{_{iP}}-\omega_i+i\epsilon}+\frac{\Lambda_i^-(p_{_{i\bot}})}{p_{_{iP}}+\omega_i-i\epsilon},\nonumber\\
&&S_i(-p'_{_2})=\frac{\Lambda_2^+(p'_{_{2\bot}})}{p'_{_{2P}}-\omega_{2f}+i\epsilon}+\frac{\Lambda_2^-(p'_{_{2\bot}})}{p'_{_{2P}}+\omega_{2f}-i\epsilon},
\end{eqnarray}
where $p_{_{iP}}\equiv \frac{p_{_i}\cdot P}{M}=\alpha_iM+J_i q_{_P}$,
$p'_{_{2P}}=\alpha'_2 P_{fP}-q_{_{fP}}=-q_{_P}-\alpha_1M+E_f$.
Substituting these propagator representations into Eq. (\ref{newa1}), performing the $q_{_P}$ integration via the residue theorem, and ignoring the contribution from negative energy wave functions, then we derive a new expression for the hadronic matrix element $\left \langle \chi_{c1}|J_\mu|B_c^+  \right \rangle$, which is given by $\mathcal{M}^B_{\mu}$:
\begin{eqnarray}\label{ampB}
\mathcal{M}^B_{\mu}=\int\frac{d^3q_{\bot}}{(2\pi)^3}Tr\left[\Lambda^+_2(p'_{2\bot})\frac{\slashed{P}_f}{M_f}\frac{M_f-\widetilde{\omega}_{1}-\widetilde{\omega}_{2}}{E_f-\omega'_{1}-\omega'_{2}}
\bar{\varphi}^{++}(q_{_{f\top}})\frac{\slashed{P}_f}{M_f}\varphi^{++}(q_{_\bot})\gamma_\mu (1 - \gamma^5)\right],
\end{eqnarray}
where $\Lambda^{+}_2(p'_{2\bot})=\frac{1}{2\omega'_{2}}\left[\frac{\slashed{P}}{M}\omega'_{2}+(-m_{2}+\slashed{p}'_{2\bot})  \right]$, $\omega'_{i}=\sqrt{{{m_i}'}^2-{p'}_{i\bot}^2}$, and \begin{eqnarray}
\bar{\varphi}^{++}(q_{_{f\top}})=\frac{\widetilde{\Lambda}^+_2(p'_{2\top})\bar{\eta}(q_{_{f\top}})
\widetilde{\Lambda}^+_1(p'_{1\top})}{M_f-\widetilde{\omega}_{1}-\widetilde{\omega}_{2}},
\end{eqnarray}
wherein the final meson's wave function is consistently evaluated in its own rest frame.
And the relation between $q_{_{f\top}}$ and $q_{_\bot}$ is
\begin{eqnarray}
q_{_{f\top}}
&=&q_{_{f\bot }}-\frac{P{_f} \cdot q_{_{f\bot }}}{M^2_f}P_f+\frac{P\cdot q_{_{f}}}{M^2}\left(P-\frac{P\cdot P_f}{M^2_f}P_f\right)\nonumber\\
&=&q_{_{\bot }}-\left(\frac{P{_f}\cdot q_{_{\bot}}+\omega'_1E_f}{M^2_f}\right)P_f+\frac{\omega'_1}{M}P.
\end{eqnarray}

\section{Wave functions of the $0^-$ and $1^{++}$ mesons}
The representation of a wave function in literature is usually given based on the $^{2S+1}L_J$, where $S$, $L$, and $J$ are the spin, orbital angular momentum, and total angular momentum of the meson, respectively. However, for a meson, orbital angular momentum $L$ is not always a good quantum number, for example, $\psi(3770)$ is a $S-D$ mixing state, and includes $L=0$ and $L=2$. We have pointed out that the expressions of $^{2S+1}L_J$, $P=(-1)^{L+1}$ and $C=(-1)^{L+S}$ are only suitable for a non-relativistic case, not for relativistic one \cite{wgl2}. For any cases, the $J^P$ is always a good quantum number for a meson. So we construct the representation of a relativistic wave function according to its $J^P$.

In general, the wave function of a pseudoscalar with $J^P=0^-$ can be expressed as a sum of 8 terms. However, under the instantaneous approximation ${P}\cdot{q_{_{\bot}}}=0$, four terms are eliminated. Then the wave function can be expressed as \cite{0-}:
\begin{eqnarray}\label{0-}
\varphi^{0^{-}}_{P}(q_{_{\bot}})=&&\left(h_{1}\slashed{P}+h_{2}M+h_{3}\slashed{q}_{{\bot}}+h_{4}\frac{\slashed{P} \slashed{q}_{{\bot}} }{M}\right)\gamma_5,
\end{eqnarray}
where the unknown radial wave function $h_i(i=1,2,3,4)$ is a function of $-q^2_{_{\bot}}$ ($=\vec{q}^2$ in the center of mass system (CMS) of $P$), and its numerical value is obtained by solving the Salpeter equation. The Salpeter equation shows us that not all $h_is$ are independent, they have the following relations \cite{0-}:
$$
h_3=\frac{h_2M(\omega_{_{2}}-\omega_{_{1}})}{m_1\omega_{_{2}}+m_2\omega_{_{1}}},\quad\quad\quad\quad
h_4=-\frac{h_1M(\omega_{_{2}}+\omega_{_{1}})}{m_1\omega_{_{2}}+m_2\omega_{_{1}}},
$$ where $\omega_{_{1}}=\sqrt{m^2_1-q^2_{_{\bot}}}$ and $\omega_{_{2}}=\sqrt{m^2_2-q^2_{_{\bot}}}$ are the energies of quarks, respectively.

It is straightforward to verify that each term of the wave function in Eq. (\ref{0-}) possesses $0^-$ quantum number. Moreover, the terms involving $h_1$ and $h_2$ correspond to $S$-waves, representing non-relativistic contributions, while those containing $h_3$ and $h_4$ correspond to $P$-waves, accounting for relativistic correction contributions  \cite{wgl2}. The positive-energy wave function of the $0^-$ meson is expressed as follows,
\begin{eqnarray}
\varphi_{0^-}^{++}=\left(B_{1}\frac{\slashed{P}}{M}+B_{2}
+B_{3}\frac{\slashed{q}_{_{\bot}}}{M}+B_{4}\frac{\slashed{P}\slashed{q}_{_{\bot}}}{M^2}\right)\gamma_{5},
\end{eqnarray}
where
$$B_1=\frac{M}{2}(h_1+h_2\frac{m_1+m_2}{\omega_{_{1}}+\omega_{_{2}}}),\quad\quad\quad B_2=\frac{\omega_{_{1}}+\omega_{_{2}}}{m_1+m_2}B_1,$$$$
B_3=\frac{M(m_2-m_1)}{m_1\omega_{_{2}}+m_2\omega_{_{1}}}B_1,\quad\quad\quad B_4=-\frac{M(\omega_{_{2}}+\omega_{_{1}})}{m_1\omega_{_{2}}+m_2\omega_{_{1}}}B_1.
$$

The positive-energy wave function of the $1^{++}$ state can be written as \cite{1++}:
\begin{eqnarray}\label{1++}
\varphi_{1^{++}}^{++}(q_{_{f\top}})=i\varepsilon_{\mu\nu\alpha\beta}\frac{P_f^\mu}{M_f}q_{_{f\top}}^\nu\epsilon^\alpha\gamma^\beta\left(F_{1}
+F_{2}\frac{\slashed{P}_f}{M_f}+F_{3}\frac{\slashed{P}_f\slashed{q}_{_{f\top}}}{M^2_f}\right),
\end{eqnarray}
where $\epsilon^\alpha$ denotes the polarization vector of the meson, $\varepsilon_{\mu\nu\alpha\beta}$ is the Levi-Civitatensor. From Eq.~(\ref{1++}), it is evident that the wave function of the $1^{++}$ state is a $P$-$D$ mixed state. The parameters $F_1$ and $F_2$ represent the non-relativistic terms, which correspond to the $P$-wave, while $F_3$ provides the relativistic correction term corresponding to the $D$-wave  \cite{wgl2}. The explicit forms of $F_1$, $F_2$, and $F_3$ can be written as:
\begin{eqnarray}
&&F_1= \frac{1}{2}\left[f_1+\frac{\widetilde{\omega}_1+\widetilde{\omega}_2}{m_1+m_2}f_2\right],           ~~~~~~~~~~~~~F_2=- \frac{1}{2}\left[\frac{m_1+m_2}{\widetilde{\omega}_1+\widetilde{\omega}_2}f_1+f_2\right],   \nonumber\\
&&F_3=-\frac{M_f(m_1+m_2)}{m_2 \widetilde{\omega}_1+m_1\widetilde{\omega}_2}F_1,             \nonumber
\end{eqnarray}
where $f_i(i=1,2)$ denote the radial wave functions of the $1^{++}$ state, and they depend on $q_{_{f\top}}^2$.

The detailed procedure for solving the complete Salpeter equations for pseudoscalar and vector mesons to derive the radial wave functions is not presented here. Interested readers may consult Refs.~\cite{0-} and~\cite{1++} for further details. The interaction potential employed in this work is displayed in Appendix~A.

\section{Results and Discussions}

In our calculations, we adopt the following parameters: $m_b$=4.96 \rm{GeV}, $m_c$=1.62 \rm{GeV}. The original mass of $\chi_{c1}(2P)$ we predicted is 3928.7 MeV, and is moved to 3871.6 MeV by varying the free parameter $V_0$ (see appendix), and the mass of $\chi_{c1}(3P)$ is 4228.8 MeV from our model. The masses of other mesons and leptons, the CKM matrix elements, and other relevant values are same as those listed in the Particle Data Group (PDG) \cite{1P5}.

\subsection{Semileptonic Decays}
\begin{figure}[!htb]
\begin{minipage}[c]{1\textwidth}
\includegraphics[width=4in]{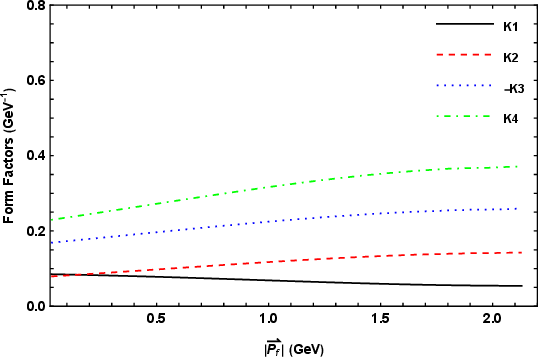}
\end{minipage}
\caption{Form Factors of $B^{+}_c\rightarrow \chi_{c1}(1P)e^+\nu_{e}$. }\label{f1}
\end{figure}

\begin{figure}[!htb]
\begin{minipage}[c]{1\textwidth}
\includegraphics[width=4in]{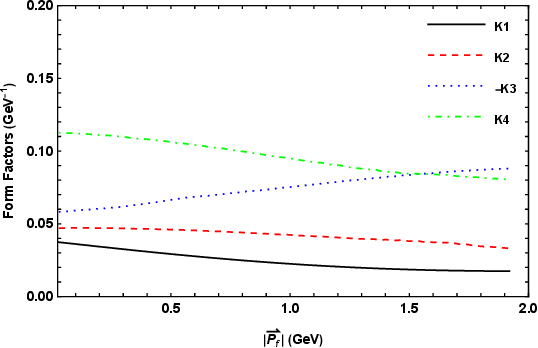}
\end{minipage}
\caption{Form Factors of $B^{+}_c\rightarrow \chi_{c1}(2P) e^+\nu_{e}. $ }\label{f2}
\end{figure}

Using the new amplitude calculation formula $\mathcal{M}^B_{\mu}$ in Eq. (\ref{ampB}), we compute the hadronic transition matrix elements for $B_c\to\chi_{c1}(nP)$. The obtained form factors defined in Eq. (\ref{transi}) for $B_c\to\chi_{c1}(1P)$ and $B_c\to\chi_{c1}(2P)$ are plotted in Figs. \ref{f1} and \ref{f2}. As can be seen from the figures, the form factors for the transition $B_c\to\chi_{c1}(1P)$ are significantly larger than those for the corresponding $B_c\to\chi_{c1}(2P)$. This is because the $\chi_{c1}(2P)$ wave functions contain nodes, and the contributions from the wave function segments before and after the nodes cancel each other, leading to a marked reduction in the results. This indicates that the decay width for the $B_c\to\chi_{c1}(2P)$ process is much smaller than that for the corresponding $B_c\to\chi_{c1}(1P)$ process.

\begin{figure}[!htb]
\begin{minipage}[c]{1\textwidth}
\includegraphics[width=4in]{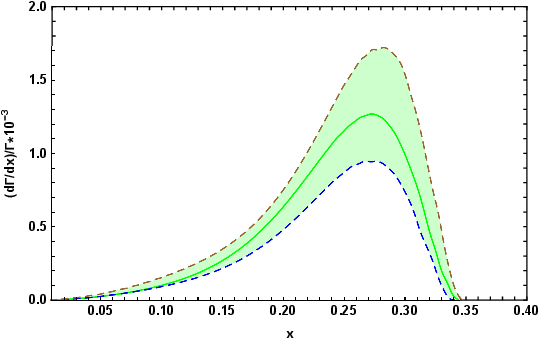}
\end{minipage}
\caption{Differential decay width $\frac{1}{\Gamma}\frac{d\Gamma}{dx}$ of $B^{+}_c\rightarrow \chi_{c1}(1P) e^+\nu_{e}$. }\label{f3}
\end{figure}

\begin{figure}[!htb]
\begin{minipage}[c]{1\textwidth}
\includegraphics[width=4in]{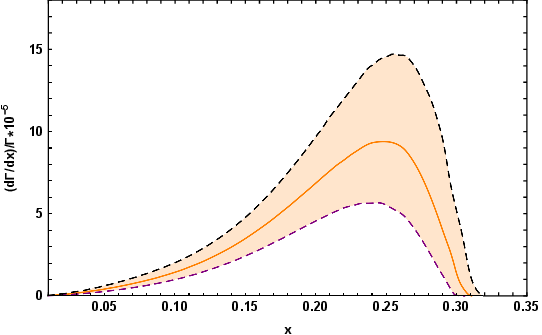}
\end{minipage}
\caption{Differential decay width $\frac{1}{\Gamma}\frac{d\Gamma}{dx}$ of $B^{+}_c\rightarrow \chi_{c1}(2P) e^+\nu_{e}$. }\label{f4}
\end{figure}

With the hadronic transition form factors available, the decay width can be calculated directly. We present in Figs. \ref{f3} and \ref{f4} the differential decay width $\frac{1}{\Gamma}\frac{d\Gamma}{dx}$ of processes $B^{+}_c\rightarrow \chi_{c1}(1P)e^+\nu_{e}$ and $B^{+}_c\rightarrow \chi_{c1}(2P)e^+\nu_{e}$, where, in addition to the central values, we also evaluate the theoretical uncertainties. The errors are obtained by randomly varying the theoretical input parameters, including the quark masses and other potential-model parameters, by $\pm5\%$.

In Table \ref{biaoyi}, we present the decay widths for the process $B^{+}_c\rightarrow \chi_{c1}(nP) \ell^+\nu_{\ell}$ (where $n=1,2,3$ and $\ell=e,\tau$) calculated using the amplitude $\mathcal{M}^B_{\mu}$. We also list several other theoretical results in the table for comparison, among which the results from Refs. \cite{19,WANG} are obtained by us using the amplitude formula $\mathcal{M}^A_{\mu}$. 
As mentioned above, our new amplitude formula $\mathcal{M}^B_{\mu}$ is superior to formula $\mathcal{M}^A_{\mu}$ in the large-recoil region. This is confirmed by the results; for example, in processes $B^{+}_c\rightarrow \chi_{c1}(1P) e^+\nu_{e}$ and $B^{+}_c\rightarrow \chi_{c1}(2P) e^+\nu_{e}$, the mass differences between the initial- and final-state particles are 2.76 GeV and 2.40 GeV, respectively, indicating that the recoil of the final state is relatively large. This leads to a significant difference in the decay widths calculated by the two methods, namely $12.9\times 10^{-16}$ GeV vs. $15.2\times 10^{-16}$ GeV and $0.96\times 10^{-16}$ GeV vs. $1.53\times 10^{-16}$ GeV, respectively. In contrast, for process $B^{+}_c\rightarrow \chi_{c1}(1P) \tau^+\nu_{\tau}$, the mass difference is 0.987 GeV, which is smaller than those of the former two processes, and the recoil effect is consequently smaller. The results are $1.30\times 10^{-16}$ GeV vs. $1.40\times 10^{-16}$ GeV, which are relatively close to each other, thus verifying the conclusion that the recoil effect is smaller in this case.

\begin{table}[!htb]
\caption{Semileptonic decay widths ($10^{-16}$ GeV) of $B_c$ to $\chi_{c1}(nP)$ and the ratio $\mathcal{R}_{\chi_{c1}(nP)}$.}
\begin{tabular}{ccccccccc}

  \hline\hline
  process  &~~This work  &~~\cite{19,WANG}  &~~\cite{DE}  &~~\cite{Akan}  &~~\cite{EH}       &~~\cite{MAI}&~~\cite{ymwang}  &~~\cite{zhangzq3}\\ \hline
  $B_c^+\rightarrow \chi_{c1}(1P)e^+\nu_{e}$  & $12.9^{+5.2}_{-3.8}$   &$15.2^{+4.5}_{-4.5}$   & $11.8$ & $11$  & $9.4^{+0.5}_{-0.3}$   &14.0&-&-\\
  $B_c^+\rightarrow \chi_{c1}(1P)\tau^+\nu_{\tau}$ & $1.30^{+1.32}_{-0.72}$   &$1.40^{+0.10}_{-0.10}$   & $1.3$ & $0.82$   & 1.0                  &1.7&-&-\\

   $\mathcal{R}_{\chi_{c1}(1P)}$& $0.100^{+0.045}_{-0.036}$   & $0.092$  & $0.11$   & $0.075$  &  $0.11$  & $0.12$ &-&-\\
  \hline
  $B_c^+\rightarrow \chi_{c1}(3872)e^+\nu_{e}$  & $0.96^{+0.61}_{-0.42}$   & 1.53           & $1.2$  & $1.9$     & -      & -&$97^{+65}_{-62}$&$12.03^{+0.97}_{-1.65}$\\
  $B_c^+\rightarrow \chi_{c1}(3872)\tau^+\nu_{\tau}$ & $0.024^{+0.052}_{-0.019}$   & -            & $0.056$  & $0.027$     & -      & -&$4.6^{+3.2}_{-3.2}$&$0.48^{+0.06}_{-0.06}$\\

   $\mathcal{R}_{\chi_{c1}(2P)}$& $0.025^{+0.024}_{-0.016}$   & -    & $0.047$   & $0.014$  &  -    & -&$0.047$&$0.040^{+0.006}_{-0.006}$\\
  \hline
  $B_c^+\rightarrow \chi_{c1}(3P)e^+\nu_{e}$  & $0.047^{+0.038}_{-0.024}$   & 0.129            &  -  &0.4   &-     & -&-&-\\
  \hline\hline
\end{tabular}\\\label{biaoyi}
\end{table}

As can be seen from the table, for the process $B_c^+ \to \chi_{c1}(1P) \ell^+ \nu_\ell$, our new result of $(12.9^{+5.2}_{-3.8})\times10^{-16}$ GeV is in good agreement with those of Refs. \cite{19,WANG,Akan,EH,MAI}. In particular, it agrees very well with the value of $11.8\times10^{-16}$ GeV from Ref. \cite{DE} and $14\times10^{-16}$ GeV from Ref. \cite{MAI}, both of which also employ the relativistic quark model. To investigate lepton universality, we also calculate the ratio of the decay widths for processes with $\tau$ and electron as the final-state particles, respectively, defined as $\mathcal{R}_{\chi_{c1}(nP)}$:
\begin{eqnarray}
\mathcal{R}_{\chi_{c1}(nP)}=\frac{\Gamma[B^+_c \to \chi_{c1}(nP)\tau^+\nu_{\tau}]}{\Gamma[B^+_c \to \chi_{c1}(nP)e^+\nu_{e}]}.
\end{eqnarray}
The results for $\mathcal{R}_{\chi_{c1}(nP)}$ ($n=1,2$) are also presented in Table \ref{biaoyi}. Our central value, $\mathcal{R}_{\chi_{c1}(1P)}=0.100$, is in very good agreement with the results of Refs. \cite{19,WANG,DE,EH}. 

For the $B_c^+ \to \chi_{c1}(3872) e^+ \nu_e$ channel, where $\chi_{c1}(3872)$ is treated as the conventional charmonium $\chi_{c1}(2P)$, we obtain a decay width of $(0.96^{+0.61}_{-0.42}) \times 10^{-16} \, \text{GeV}$. Compared with the result for the $1P$ final state, the decay width decreases by about one order of magnitude, which is mainly due to the presence of a node in the wave function of the $2P$ state, and the contributions from the wave function on either side of the node cancel strongly, resulting in a very small width. Although our result is close to those of Refs. \cite{19,WANG,DE,Akan}, it is much smaller than the values of Refs. \cite{ymwang,zhangzq3}, indicating that the situation for this decay channel is far more complicated than we had expected, and thus deserves more attention and further in-depth studies.

For $B_c^+ \to \chi_{c1}(3872) \tau^+ \nu_\tau$, we obtain a decay width of $(2.4^{+5.2}_{-1.9}) \times 10^{-18} \, \text{GeV}$, which is smaller than that of the decay channel $B_c^+ \to \chi_{c1}(1P) e^+ \nu_e$ by almost three orders of magnitude. This results from the nodal structure of the $2P$ state combined with a strong suppression from the phase space. 
For the $B_c^+ \to \chi_{c1}(3P) e^+ \nu_e$
channel, we also obtain a very small decay width of $(4.7^{+3.8}_{-2.4}) \times 10^{-18} \, \text{GeV}$. This suppression is mainly due to the more complex two-node structure in the wave function of the $3P$ state, reflecting that the increasing number of nodes in the wave functions leads to a continuously decreasing overlap integral as the radial quantum number $n$ increases. 

Although the new amplitude formula $\mathcal{M}^B_{\mu}$ has an advantage over our previously used formula $\mathcal{M}^A_{\mu}$ in the large-recoil region, as it can better handle the transition behavior at large recoil, we note that formula $\mathcal{M}^B_{\mu}$ also has a drawback: due to its greater complexity, it introduces larger uncertainties.

\subsection{Nonleptonic Decays}
There are some parameters appeared only in nonleptonic decays, and they are chosen as $a_1$=1.14, $f_{\pi}=0.130$ GeV, $f_{\rho}=0.2085$ GeV, $f_{K}=0.156$ GeV, $f_{K^*}=0.217$ GeV, $f_{D^{+}}=0.214$ GeV,
$f_{D^{*+}}=0.27$  GeV, $f_{D_s^{+}}=0.251$ GeV, and $f_{D_s^{*+}}=0.3$ GeV.

In terms of nonleptonic decays, we apply the amplitude formula $\mathcal{M}^B_{\mu}$ to calculate some color-favored decay channels, and present the results in Table \ref{tab2}. We also list some other theoretical results for comparison. As can be seen from the comparison, the results for nonleptonic processes differ from those for semileptonic processes. In the semileptonic decays, many theoretical results are close to each other, but in the nonleptonic decays, the agreement among different results is poor. Our current results calculated using $\mathcal{M}^B_{\mu}$ are in relatively good agreement with those obtained using formula $\mathcal{M}^A_{\mu}$ \cite{19,WANG}, and are partially comparable to those in Ref. \cite{DE}, but show discrepancies from the results of other theoretical models. 

\begin{table}[!htb]
\caption{Nonleptonic decay widths ($10^{-17}$ GeV) of $B_c\to\chi_{c1}(nP)X$.}
\begin{tabular}{ccccccc}
  \hline \hline
  process                 &~~~~This work      &~~~~\cite{zhangzq2}     &~~~~\cite{DE}   &~~~~\cite{W.wang} &~~~~\cite{19,WANG}      &~~~~\cite{EH}     \\ \hline
  $B_c^+\rightarrow \chi_{c1}(1P)\pi^+$    &~~~~ $3.56^{+0.55}_{-0.56}$ &~~~~$17^{+0}_{-0}$          &~~~~ $28$  &~~~~-    &~~~~$3.1\pm0.3$    &~~~~ 0.19   \\
  $B_c^+\rightarrow \chi_{c1}(1P)\rho^{+}$ &~~~~ $28.4^{+2.2}_{-2.5}$   &~~~~$55.6^{+2.6}_{-2.6}$         &~~~~ $21$  &~~~~-   &~~~~$33\pm3$    &~~~~14       \\
  $B_c^+\rightarrow \chi_{c1}(1P) K^+$  &~~~~ $0.265^{+0.030}_{-0.032}$  &~~~~$1.3^{+0.0}_{-0.0}$      &~~~~ $2.1$ &~~~~-    &~~~~$0.23\pm0.03$    &~~~~ 0.016       \\
  $B_c^+\rightarrow \chi_{c1}(1P)K^{*+}$  &~~~~ $1.96^{+0.16}_{-0.16}$   &~~~~$3.5^{+1.3}_{-1.3}$       &~~~~ $1.4$ &~~~~-    &~~~~$2.5\pm0.1$    &~~~~ 1.0       \\  \hline
  $B_c^+\rightarrow \chi_{c1}(3872)\pi^+$   &~~~~ $0.81^{+0.12}_{-0.16}$ &~~~~$32.3^{+5.2}_{-2.6}$          &~~~~ $1.4$ &~~$24^{+17}_{-17}$    &~~~~-    &~~~~ -       \\
  $B_c^+\rightarrow \chi_{c1}(3872)\rho^{+}$ &~~~~ $5.20^{+1.31}_{-1.27}$    &~~~~$81.3^{+9.0}_{-15.5}$      &~~~~ $2.0$ &~~$59^{+29}_{-23}$     &~~~~-    &~~~~ -       \\
  $B_c^+\rightarrow \chi_{c1}(3872)K^+$  &~~~~ $0.058^{+0.010}_{-0.012}$    &~~~~$2.6^{+0.1}_{-0.1}$    &~~~~$0.12$ &~~$1.9^{+1.3}_{-1.4}$    &~~~~-    &~~~~ -       \\
  $B_c^+\rightarrow \chi_{c1}(3872)K^{*+}$   &~~~~ $0.342^{+0.067}_{-0.077}$    &~~~~$4.6^{+0.8}_{-0.6}$     &~~~~$0.14$ &~~$3.4^{+2.3}_{-2.3}$    &~~~~-    &~~~~ -      \\
  $B_c^+\rightarrow \chi_{c1}(3872)D^+$  &~~~~ $0.028^{+0.018}_{-0.016}$      &~~~~$4.3^{+0.5}_{-0.4}$   &~~~~-  &~~~~-    &~~~~-    &~~~~ -       \\
  $B_c^+\rightarrow \chi_{c1}(3872)D^{*+}$  &~~~~ $0.440^{+0.305}_{-0.240}$   &~~~~$10.1^{+2.2}_{-1.8}$        &~~~~-  &~~~~-    &~~~~-    &~~~~ -      \\
  $B_c^+\rightarrow \chi_{c1}(3872)D_s^+$     &~~~~ $0.522^{+0.428}_{-0.339}$  &~~~~$129^{+41}_{-35}$     &~~~~-  &~~~~-    &~~~~-    &~~~~ -        \\
  $B_c^+\rightarrow \chi_{c1}(3872)D_s^{*+}$    &~~~~ $7.41^{+5.87}_{-3.93}$    &~~~~$230^{+28}_{-23}$       &~~~~-  &~~~~ -   &~~~~-    &~~~~ -        \\  \hline
  $B_c^+\rightarrow \chi_{c1}(3P)\pi^+$   &~~~~ $0.091^{+0.032}_{-0.029}$   &~~~~-       &~~~~-  &~~~~-    &~~~~-    &~~~~ -       \\
  $B_c^+\rightarrow \chi_{c1}(3P)\rho^+$    &~~~~ $0.407^{+0.176}_{-0.149}$    &~~~~-    &~~~~-  &~~~~-    &~~~~-    &~~~~ -       \\
  $B_c^+\rightarrow \chi_{c1}(3P)K^+$      &~~~~ $0.0063^{+0.0023}_{-0.0021}$ &~~~~-   &~~~~-  &~~~~-    &~~~~-    &~~~~ -       \\
  $B_c^+\rightarrow \chi_{c1}(3P)K^{*+}$ &~~~~ $0.0255^{+0.0117}_{-0.0094}$      &~~~~-    &~~~~-  &~~~~-    &~~~~-    &~~~~ -       \\
  \hline \hline
\end{tabular}\label{tab2}
\end{table}

Since in two-body nonleptonic decays the phase space is fixed, corresponding to a single point in the form factor. The fact that the semileptonic results agree well while the nonleptonic results show large discrepancies indicates that different models yield significantly different results at this phase-space point, that is, different models differ greatly in their treatment of recoil effects.  For example, comparing the $B_c^+ \to \chi_{c1}(1P) \pi^+$ and $B_c^+ \to \chi_{c1}(1P) \rho^+$ (or $B_c^+ \to \chi_{c1}(1P) K^{*+}$) decays, the mass of the $\rho^+$ (or $K^{*+}$) meson is larger than that of the $\pi^+$; therefore, the recoil in the former process is larger than in the latter, meaning that the phase-space point corresponding to the latter is closer to the zero-recoil point. In general, form factor calculations at points close to the zero-recoil point are relatively more accurate in most models. The comparison of results in Table \ref{tab2} confirms this conclusion: for the
$B_c^+ \to \chi_{c1}(1P) \rho^+$ process, the theoretical values from various models show relatively good agreement, whereas for 
$B_c^+ \to \chi_{c1}(1P) \pi^+$, the agreement is poor.

Regarding the nonleptonic decays of $B_c^+ \to \chi_{c1}(3872) X$, we note that their decay widths are smaller than those of the corresponding $B_c^+ \to \chi_{c1}(1P) X$ processes by factors of $4.4$ to $5.7$. This again primarily reflects the suppression of the $2P$ states due to the nodal structure of the wave functions. In addition to the $X=\pi,~\rho,~K,~K^*$ channels, we have also calculated processes with final-state $X$ being $D$, $D^*$, $D_s$, $D^*_s$. In particular, the 
$B_c^+ \to \chi_{c1}(3872) D^*_s$ channel has a relatively large width of $(7.41^{+5.87}_{-3.93})\times 10^{-17}$ GeV. For the decays of $B_c^+ \to \chi_{c1}(3P) X$, since the $3P$ wave functions have two nodes, these processes are suppressed even more. Of course, phase-space suppression also plays a role; for example, the channels with final state $X$ being $D$, $D^*$, $D_s$, and $D^*_s$ are kinematically forbidden.

\subsection{Possible experimental searches}

We note that the LHCb Collaboration has already attempted to search for the  particle $\chi_{c1}(3872)$ in $B_c$ decays \cite{11}. They measured the ratio (Eq. (\ref{eq1})):
$$
\mathcal{R}_{\psi(2S)}^{\chi_{c1}(3872)} = \frac{\mathcal{B}_{B_c^+ \to \chi_{c1}(3872)\pi^+}}{\mathcal{B}_{B_c^+ \to \psi(2S)\pi^+}} \times \frac{\mathcal{B}_{\chi_{c1}(3872) \to J/\psi \pi^+\pi^-}}{\mathcal{B}_{\psi(2S) \to J/\psi \pi^+\pi^-}}. 
$$
However, they did not observe a definite value, but instead set an upper limit of 
$\mathcal{R}_{\psi(2S)}^{\chi_{c1}(3872)}< 0.05 \, (0.06)$. 

We can provide a theoretical prediction for this ratio $\mathcal{R}_{\psi(2S)}^{\chi_{c1}(3872)}$. For the branching fractions $\mathcal{B}_{\chi_{c1}(3872) \to J/\psi \pi^+\pi^-}$ and $\mathcal{B}_{\psi(2S) \to J/\psi \pi^+\pi^-}$, we adopt the experimental data from PDG \cite{1P5}. The branching ratio 
$\mathcal{B}_{B_c^+ \to \psi(2S)\pi^+}=0.0266\%$ is taken from our previous calculation using amplitude formula $\mathcal{M}^A_{\mu}$ in Ref. \cite{fuhf2}. In that paper, we obtained the ratios $\frac{\mathcal{B}_{B_c^+ \to \psi(2S)\pi^+}}{\mathcal{B}_{B_c^+ \to \psi(1S)\pi^+}}=0.240^{+0.023}_{-0.040}$ and $\frac{\mathcal{B}_{B_c^+ \to \psi(1S)K^+}}{\mathcal{B}_{B_c^+ \to \psi(1S)\pi^+}}=0.0763$, which are in good agreement with the experimental data of $0.254\pm0.018\pm0.006$ and $0.079\pm0.007\pm0.003$ \cite{1P5}, demonstrating the feasibility of our approach. 
​The value of $\mathcal{B}_{B_c^+ \to \chi_{c1}(3872)\pi^+}=6.31\times10^{-6}$ is taken from the results of this work. We thus obtain $\mathcal{R}_{\psi(2S)}^{\chi_{c1}(3872)} = 0.0024$, which is consistent with the experimental upper limit $\mathcal{R}_{\psi(2S)}^{\chi_{c1}(3872)}< 0.05 \, (0.06)$. Our results also indicate that the current accumulated $B_c$ event sample is insufficient to observe the $B_c\to \chi_{c1}(3872)\pi$ process. To detect this decay, we estimate that about 20 times more $B_c$  events would be needed.

\begin{table}[!htb]
\caption{Branching ratios of $B_c\to\chi_{c1}(nP)$ decays.}
\begin{tabular}{cccc}

  \hline\hline
  process          &~~~~$\mathcal{B}$  &~~~~ process            &~~~~$\mathcal{B}$\\ \hline
  $B_c^+\rightarrow \chi_{c1}(1P)e^+\nu_{e}$    &~~$(1.00^{+0.40}_{-0.30}) \times 10^{-3}$&~~~~~~$B_c^+\rightarrow \chi_{c1}(1P)\tau^+\tau_{\ell}$   &~~$(1.01^{+1.02}_{-0.56}) \times 10^{-4}$\\ 
  $B_c^+\rightarrow \chi_{c1}(3872)e^+\nu_{e}$    &~~$(7.4^{+4.7}_{-3.3}) \times 10^{-5}$&~~~~~~$B_c^+\rightarrow \chi_{c1}(3872)\tau^+\tau_{\ell}$   &~~$(1.84^{+4.05}_{-1.46}) \times 10^{-6}$\\ 
  $B_c^+\rightarrow \chi_{c1}(3P)e^+\nu_{e}$    &~~$(3.64^{+2.91}_{-1.88}) \times 10^{-6}$&~~~~&\\
  \hline

 $B_c^+\rightarrow \chi_{c1}(1P)\pi^+$     &~ $(2.76^{+0.42}_{-0.44} ) \times10^{-5}$   &~~~~~~  $B_c^+\rightarrow \chi_{c1}(1P)\rho^{+}$    &~ $(2.21^{+0.17}_{-0.20} ) \times 10^{-4}$ \\

 $B_c^+\rightarrow \chi_{c1}(1P) K^+$   &~ $(2.05^{+0.24}_{-0.25} ) \times 10^{-6}$ &~~~~~~ $B_c^+\rightarrow \chi_{c1}(1P)K^{*+}$     &~ $(1.52^{+0.12}_{-0.12} ) \times 10^{-5}$       \\ \hline

 $B_c^+\rightarrow \chi_{c1}(3872)\pi^+$   &~ $(6.31^{+0.97}_{-1.24} ) \times 10^{-6}$ &~~~~~~  $B_c^+\rightarrow \chi_{c1}(3872)\rho^{+}$        &~ $(4.02^{+1.01}_{-0.99} ) \times 10^{-5}$       \\

 $B_c^+\rightarrow \chi_{c1}(3872)K^+$              &~ $(4.50^{+0.75}_{-0.92} ) \times 10^{-7}$  &~~~~~~$B_c^+\rightarrow \chi_{c1}(3872)K^{*+}$           &~ $(2.65^{+0.52}_{-0.60} ) \times 10^{-6}$      \\

 $B_c^+\rightarrow \chi_{c1}(3872)D^+$              &~ $(2.21^{+1.44}_{-1.24} ) \times 10^{-7}$    &~~~~~~
  $B_c^+\rightarrow \chi_{c1}(3872)D^{*+}$           &~ $(3.41^{+2.36}_{-1.86} ) \times 10^{-6}$      \\

   $B_c^+\rightarrow \chi_{c1}(3872)D_s^+$          &~ $(4.05^{+3.32}_{-2.62}) \times 10^{-6}$      &~~~~~~
  $B_c^+\rightarrow \chi_{c1}(3872)D_s^{*+}$        &~ $(5.75^{+4.55}_{-3.05} ) \times 10^{-5}$        \\ \hline

   $B_c^+\rightarrow \chi_{c1}(3P)\pi^+$           &~ $(7.03^{+2.49}_{-2.26} ) \times 10^{-7}$    &~~~~~~
  $B_c^+\rightarrow \chi_{c1}(3P)\rho^+$           &~ $(3.16^{+1.36}_{-1.15} ) \times 10^{-6}$      \\
 $B_c^+\rightarrow \chi_{c1}(3P)K^+$              &~ $(4.85^{+1.80}_{-1.65} ) \times 10^{-8}$      &~~~~~~
  $B_c^+\rightarrow \chi_{c1}(3P)K^{*+}$           &~ $(1.97^{+0.91}_{-0.72} ) \times 10^{-7}$

\\  \hline\hline
\end{tabular}\\\label{tab3}
\end{table}

To compare the experimental detectability of various decay channels, we present the branching fractions for $B_c\to\chi_{c1}(nP)$ in Table \ref{tab3}. As can be seen from the table, 
$\mathcal{B}_{B_c^+ \to \chi_{c1}(3872)\pi^+}=6.31^{+0.97}_{-1.24}\times 10^{-6}$, which is smaller than 
$\mathcal{B}_{B_c^+ \to \chi_{c1}(3872)\rho^+}=4.02^{+1.01}_{-0.99}\times 10^{-5}$, $\mathcal{B}_{B_c^+ \to \chi_{c1}(3872)D^{*+}_s}=5.75^{+4.55}_{-3.05}\times 10^{-5}$, and 
$\mathcal{B}_{B_c^+ \to \chi_{c1}(3872)e^+\nu_{e}}=7.4^{+4.7}_{-3.3}\times 10^{-5}$. 
Since neither the $\rho$ nor the $D^{*+}_s$ is directly accessible to experimental probes, these decay modes are disfavored. Therefore, the experimental identification of $B_c^+ \to \chi_{c1}(3872)e^+\nu_{e}$
is considerably more tractable than that of $B_c^+ \to \chi_{c1}(3872)\pi^+$. At LHCb, the muon channel $B_c^+ \to \chi_{c1}(3872)\mu^+\nu_{\mu}$  is the most sensible choice, requiring only about twice the current $B_c$ sample size for observation.

In Table \ref{ratio}, we present the ratio $\mathcal{R}(X)$, defined as the ratio of the decay rate of the nonleptonic decay $\mathcal{B}_{B_c^+ \to \chi_{c1}(3872)X}$ to the semileptonic decay $\mathcal{B}_{B_c^+ \to \chi_{c1}(3872)e^+\nu_e}$, i.e.,
$$\mathcal{R}(X)\equiv \frac{\mathcal{B}_{B_c^+ \to \chi_{c1}(3872)X}}{\mathcal{B}_{B_c^+ \to \chi_{c1}(3872)e^+\nu_{e}}},$$
in order to provide a clear comparison of the branching fractions among various decay channels. As mentioned above, this ratio may be universal and relatively reliable \cite{qzhao}. As can be seen from the table, for all nonleptonic 
$B_c^+ \to \chi_{c1}(3872)$ decays, the branching fractions are smaller than that of the semileptonic process $B_c^+ \to \chi_{c1}(3872)e^+\nu_{e}$. In particular, for $B_c^+ \to \chi_{c1}(3872)\pi^+$, which the LHCb experiment has attempted to search for, we obtain $\mathcal{R}(\pi^+)=0.085$, meaning that its branching fraction is about 12 times smaller than that of 
$B_c^+ \to \chi_{c1}(3872)e^+\nu_{e}$, making it rather challenging to observe.

\begin{table}[ht]
\centering
\caption{Ratio $\mathcal{R}(X)=\mathcal{B}_{B_c^+ \to \chi_{c1}(3872)X} / \mathcal{B}_{B_c^+ \to \chi_{c1}(3872)e^+\nu_e}$.} 
\setlength{\tabcolsep}{6pt}
\renewcommand{\arraystretch}{1}
\begin{tabular}{|cc|cc|cc|cc|}
\hline\hline
$\mathcal{R}(\pi^+)$ & 0.085 & $\mathcal{R}(\rho^{+})$ &  0.543 &$\mathcal{R}(K^+)$& 0.0061  
&$\mathcal{R}(K^{*+})$ & 0.036   \\
 $\mathcal{R}(D^+ )$ & 0.0030 &$\mathcal{R}(D^{*+} )$& 0.046 &
$\mathcal{R}(D_s^+)$ & 0.055   & $\mathcal{R}(D_s^{*+})$ & 0.777 
\\ \hline\hline
\end{tabular}\label{ratio}
\end{table}

\section{Summary}
We investigate the semileptonic weak decays $B_c^+ \to \chi_{c1}(nP) \ell^+ \nu_\ell$
($n$=1, 2, 3, $\ell=e,~\tau$) and the corresponding color-favored nonleptonic channels $B_c^+ \to \chi_{c1}(nP) X^+$ ($X=\pi,~\rho,~K,$ etc.), treating $\chi_{c1}(3872)$ as the conventional charmonium state $\chi_{c1}(2P)$. Motivated by recent LHCb measurements searching for $B_c^+ \to \chi_{c1}(3872) \pi^+$, we revisit these transitions with improved transition amplitude, which is evaluated in the Bethe–Salpeter framework under the instantaneous approximation, employing Salpeter wave functions constructed according to $J^P$ quantum numbers rather than the traditional $^{2S+1}L_J$ scheme. A covariant amplitude $\mathcal{M}^B_{\mu}$ is proposed to properly account for large-recoil kinematics and relativistic corrections, which are especially significant for excited charmonium states such as $\chi_{c1}(2P)$ and $\chi_{c1}(3P)$. 

Numerical results show that the decay widths decrease markedly with increasing radial quantum number due to nodal structures in the wave functions.
Although sizable branching fractions are predicted for $B_c^+ \to \chi_{c1}(3872) \rho^+$ and $B_c^+ \to \chi_{c1}(3872) D^{*+}_s$, while the $\pi$ channel $B_c^+ \to \chi_{c1}(3872) \pi^+$ is relatively suppressed. We further predict the ratio $\mathcal{R}_{\psi(2S)}^{\chi_{c1}(3872)} = 0.0024$, consistent with the current LHCb upper limit $< 0.05 \, (0.06)$ and showing that approximately 20 times the current $B_c$ data sample would be required for a definitive observation.
We suggest that detection of $B_c^+ \to \chi_{c1}(3872) \mu^+ \nu_\mu$ is feasible with approximately twice the existing $B_c$ data sample in LHCb. Our results provide important theoretical inputs for ongoing experimental efforts to probe the internal structure of $\chi_{c1}(3872)$ via $B_c$ weak decays.

\vspace{0.7cm} {\bf Acknowledgments}

This work is supported by the National Natural Science Foundation of China under the Grants No. 12575097, No. 12375085, and No. 12365013. W. Li is also supported by Natural Science Foundation of Hebei province under the Grant No. A2025204003 and Hebei Agricultural University introduced talent research special project (No. YJ2024038). T. Wang is also supported by the Fundamental Research Funds for the Central Universities (2023FRFK06009).

\section{APPENDIX A}
Since the wave function is relativistic, to avoid double counting, the interaction potential must be non-relativistic when solving the Salpeter equation.
We adopt the Cornell potential, which consists of a linear confining potential plus a Coulomb potential arising from single-gluon exchange
\begin{equation}
V(r)= \lambda r+V_0- \gamma_0\otimes\gamma^0\frac{4}{3}\frac{\alpha_s}{r},\label{potential}
\end{equation}
where $\lambda=0.21$ $\rm GeV^2$ is the string tension, $V_0$ is a free constant appearing in potential model to fit data, $\alpha_s$ is the running coupling constant. To avoid divergence in momentum space and account for the screening effect, a factor $e^{-\alpha r}$ ($\alpha=0.06$ $\rm GeV$) is added \cite{Ding,Chao}:
\begin{equation}
V(r)= \frac{\lambda}{\alpha}(1-e^{-\alpha
r})+V_0-\gamma_0\otimes\gamma^0\frac{4}{3}\frac{\alpha_s}{r}e^{-\alpha
r}.\label{potential1}
\end{equation}
And its representation in momentum space is \cite{0-}:
\begin{eqnarray}\label{potential2}
V(\vec{q})=&&V_{s}(\vec{q})+V_{v}(\vec{q})\gamma_{0}\otimes\gamma^{0},
\nonumber\\
V_{s}(\vec{q})=&&-(\frac{\lambda}{\alpha}+V_{0})\delta^{3}(\vec{q})+\frac{\lambda}{\pi^{2}}\frac{1}{(\vec{q}^{2}+\alpha^{2})^{2}},
\nonumber\\
V_{v}(\vec{q})=&&-\frac{2}{3\pi^{2}}\frac{\alpha_{s}(\vec{q})}{(\vec{q}^{2}+\alpha^{2})},
\nonumber\\
\alpha_{s}(\vec{q})=&&\frac{12\pi}{27}\frac{1}{log(a+\frac{\vec{q}^{2}}{\Lambda_{QCD}^{2}})},
\end{eqnarray}
where $\Lambda_{QCD}=0.27$ $\rm GeV$ is confinement energy scale and $a=e=2.7183$.

\end{document}